\documentclass[useAMS,usenatbib]{mn2e}
\def\aap{AA}

\def\apjl{ApJL}

\def\mnras{MNRAS}
\def\apj{ApJ}
\def\apjs{ApJS}

\def\aj{AJ}

\def\pasp{PASP}

\def\procspie{Proc.~SPIE}
\def\nat{Nat}

\def\ase{{\prime\prime}}

\usepackage{graphicx}
\usepackage{float}
\usepackage{amssymb}
\usepackage{amsfonts}
\usepackage{amsmath} 
\usepackage{color}
\def\aaemail{\tt aagnello@eso.org}
\def\eso{European Southern Observatory, Karl-Schwarzschild-Strasse 2, 85748 Garching bei M{\"u}nchen, DE}

\title[Outlier search, GAIA, lensed quasars and MW streams]{Quasar Lenses and Galactic Streams: Outlier Selection and GAIA Multiplet Detection} 
\author[A. Agnello]{
  Adriano Agnello$^{1}$\thanks{\aaemail}
  \medskip\\
  $^1$\eso\\
}

\begin{document}

\voffset-.6in

\date{Accepted . Received }

\pagerange{\pageref{firstpage}--\pageref{lastpage}} 

\maketitle

\label{firstpage}

\begin{abstract}
I describe two novel techniques originally devised to select strongly lensed quasar candidates in wide-field surveys. The first relies on outlier selection in optical and mid-infrared magnitude space; the second combines mid-infrared colour selection with GAIA spatial resolution, to identify multiplets of objects with quasar-like colours. Both methods have already been applied successfully to the SDSS, ATLAS and DES footprints: besides recovering known lenses from previous searches, they have led to new discoveries, including quadruply lensed quasars, which are rare within the rare-object class of quasar lenses. As a serendipitous by-product, at least four candidate Galactic streams in the South have been identified among foreground contaminants. There is considerable scope for tailoring the WISE-GAIA multiplet search to stellar-like objects, instead of quasar-like, and to automatically detect Galactic streams.
\end{abstract}

\section{Introduction}
The many virtues of strongly lensed quasars, both for astrophysics and cosmography, are hampered by the relative paucity of systems with sufficient ancillary data.
On the one hand, lensed quasars are valuable probes of cosmological distances and stellar and dark matter in z$\approx$1 galaxies \citep{cou02}: the delays between light-curves of different quasar images can be translated in one-step measurements of the Hubble constant \citep{ref64}, making a low-redshift complement to CMB experiments \citep{suy16}; lens modeling and source-reconstruction enables super-resolved studies of quasar host galaxies at z$\approx$2 \citep{din17}; and micro-lensing yields a detailed view of the source central engine  \citep{bat11,slu12,hut15} and luminous and dark mass in the deflector \citep{sch14,ogu14}.
On the other hand, a few tens of these systems are currently known and few of them are suitable for time-delay cosmography or detailed lens modeling, whence the need for larger samples. With one every $\mathcal{O}(10^4)$ quasars being strongly lensed \citep{om10}, and $\approx$0.2 quasar lenses per square degree, these are a class of rare objects to be mined in wide-field surveys. In particular, predicted quasar lens samples are predominantly doubles, with only $\approx14\%$ being highly valuable quads.

Different searches, tailored to different data-sets and surveys, have been developed to find new lensed quasars. The Cosmic Lens All Sky Survey \citep[CLASS,][]{mye03} and a parallel JVLA search \citep{kin99} targeted radio-loud objects that could be resolved in multiple components by follow-up observations; the Sloan Quasar Lens Search \citep[SQLS,][]{ogu06} and its BOSS extension \citep[BQLS,][]{mor16} targeted objects that were identified as quasars from previous fibre spectroscopy. Both CLASS and SQLS/BQLS uncovered the bright end of the lensed quasar population, and were focused on either radio or UV excess pre-selection \citep[for the challenges in observing radio-quiet lensed quasars, see e.g.][]{jac15}. In order to expand this effort to fainter systems and higher redshift, a variety of techniques \citep{mor04,agn15,wil17,ost17} have been tailored to wide-field photometric and morphological searches.

Here, I illustrate two novel search methods that I have developed and used to discover new quasar lenses and whose performance has been complementary to other searches applied over the last year. The first, described in Section 2, relies on outlier selection in the optical and mid-infrared, selecting lensed quasar targets among objects that do not `obviously' belong to classes/clusters of more common contaminants.
The second, described in Section 3, combines the good spatial resolution and depth of the GAIA mission \citep{gai16,lin16,vLe17} with a mid-infrared colour selection of quasar-like objects. Four new Galactic streams, found as a by-product of this method, are also briefly described. Given the setup of this search (tailored on quasar lenses), this discovery may be considered serendipitous, so I will outline how a similar search can be tailored on Galactic substructure. Concluding remarks are summarized in Section 4.
 Discoveries of quasar lenses from both methods, involving different collaborations, are reported elsewhere.

 In what follows, some nomenclature will be consistently used. \textit{Objects} are selected at query level from wide-field surveys, \textit{targets} are a sub-sample of objects selected based on their catalog properties, and \textit{candidates} are a sub-sample of targets further selected based on their images either via visual inspection or cutout modeling. For the sake of brevity, I designate as quasar `pairs' both physically associated quasars and chance alignments of quasars at different redshifts. The same holds for line-of-sight (LOS) quasar-star `pairs'.
 
 Wide-field surveys are abbreviated as follows: SDSS is the Sloan Digital Sky Survey \citep{aba09}; ATLAS is the VLT Survey Telescope ATLAS survey \citep{sha15}; WISE is the Wide Infrared Survey Imager \citep{wri10}; DES is the Dark Energy Survey \citep{san10}; iPTF is the intermediate Palomar Transient Factory \citep{law09} DR3; and PS1 is the first Pan-STARRS \citep{kai10,cha16} telescope data release. Throughout this paper, $griz$ magnitudes are in the AB system, and mid-infrared $W1,$ $W2,$ $W3$ magnitudes from WISE are in the Vega system.
\begin{table*} 
\centering
\begin{tabular}{lccccccccl}
\hline
$k$	&	$g-r$	&	$g-i$	&	$r-z$ &	$i-W1$ &	$W1-W2$ &	$W2$ &	$W2-W3$\\
\hline
1 & 0.15 & 0.44 & 0.48 & 4.34 & 0.89 & --- & 2.80\\
	& 0.31 & 0.37 & 0.21 & 0.32 &  0.14 & --- & 0.32\\
2 & 0.32 & 0.52 & 0.25 & 4.45 & 0.55 & 16.30 & 3.90 \\
   & 0.20 & 0.32 & 0.29 & 1.07 &  0.61 & 0.35 & 0.50\\
3 &  0.16 & 0.14 & 0.00 & 3.60 & 1.24 & 14.90 &  2.94\\
   &  0.10 & 0.13 & 0.17 & 0.15 & 0.11 & 0.47 & 0.15\\
4	&	0.95 &1.40 & 0.78 & 3.51& 0.55 & --- & 3.5 \\
    &  0.24 & 0.30 & 0.16 & 0.41& 0.40 & --- & 0.55 \\
\hline
5 & 0.35 & 0.61 &  0.61 & 4.17 & 0.82 & 14.19 & 3.12\\
   &  0.16 & 0.18 & 0.16 & 0.48 & 0.13 & 0.32 & 0.20\\
6 &  0.48 & 0.88 & 0.84 & 4.16 & 0.92 & 13.90 & 2.35\\
   &  0.25 & 0.25 & 0.12 & 0.47 & 0.21 & 0.48 & 0.36\\
\hline
7 & 0.22 & 0.25 & 0.44 & 4.50 & 0.79 & 14.20 & 3.10\\
   & 0.22 & 0.38 & 0.21 & 0.47 & 0.11 & 0.30 & 0.26\\
8 & 0.17 & 0.25 & 0.60 & 4.56 & 1.03 & 13.18 & 2.49\\
   & 0.24 & 0.39 & 0.23 & 0.46 & 0.12 & 0.53 & 0.28\\
9 & 0.15 & 0.25 & 0.32 & 3.05 & 1.00 & 14.12 & 3.20\\
   &  0.11 & 0.12 & 0.10 & 0.20 & 0.13 & 0.33 & 0.28\\
10 & 0.13 & 0.31 & 0.34 & 3.04 & 1.26 & 13.63 & 3.30\\
     &  0.09 & 0.15 & 0.10 & 0.25 & 0.09 & 0.37 & 0.23\\
\hline
\end{tabular}
\caption{Means $\mu_{k,j}$ and widths $\sqrt{\Sigma_{k,ii}}$ of the object clusters computed as in S~\ref{sect:clusters}; the association between clusters and classes is done a posteriori, and described in the text.}
\label{tab:feattab}
\end{table*}
\section{Outlier selection}

Depending on the survey image quality and depth, and lens configuration and image separation, lensed quasars have colours intermediate between those of the source quasars and those of the deflector galaxies. Likewise, they can result in groups of point-like or extended sources, or as extended objects due to blending by image processing pipelines. Due to their rarity and intermediate colours and morphologies, lensed quasar candidate samples suffer from significant contamination by more common classes  of objects. This problem can be mitigated by selecting objects that do \textit{not} have typical colours of more common contaminants.

As a \textit{training} set for the outlier-selection method, I use the 10 quasar lenses and 40 false-positives in the morphologically-selected sample of \citet{ina12}. This is because most quasar lenses are marginally deblended by the pipelines of ground-based surveys, and those with larger separation in previous searches have colours that are more typical of nearby quasars, due to the UVx and spectroscopic pre-selections that were applied there. The 10+40 morphological candidates of \citet{ina12} should then be a good guidance to a homogeneous sample of quasar-like objects with nearby companions, with different kinds of contaminants, and small enough that data mining techniques trained on it do not over-specialize, thus remaining complementary to previous searches.

As a \textit{test} set, to evaluate the method performance on a wider sample, I use a list of 149 known lenses compiled from the CASTLES\footnote{\texttt{https://www.cfa.harvard.edu/castles/}} database and SQLS full sample from SDSS-DR7 given by \citet{ina12}. In particular, I will consider the 132 known systems with DEC$>-20$ for tests on the SDSS footprint. These are the same used by \citet{wil17} to evaluate the performance of population-mixture methods on quasar lenses in different ranges of image separation.
Some of the outlier-selected targets were later found to be already known quasar lenses and pairs in the BQLS sample of \citet{mor16}.

\subsection{Setup, clusters and object classes}
\label{sect:clusters}
In order to describe different populations in colour-magnitude space, and similarly to previous work \citep{wil17}, I  consider $g-r,$ $g-i,$ $r-z,$ $i-W1,$ $W2,$ $W1-W2$ and $W2-W3,$ compressing the catalog information to a seven-dimensional \textit{feature} space. I do not use UV excess information, since it is not always available in current wide-field surveys (e.g. DES), and it is less efficient at identifying quasars at redshifts $z_{s}\gtrsim2.5,$ where Ly$\alpha$ emission exits the $u-$band.

Object pre-selection is based on their extended morphology
 and some loose colour-magnitude requirements. As a morphological pre-selection, I concentrate on objects that have\texttt{psf{\_}r}-\texttt{mod{\_}r}$>0.075$ and \texttt{psf{\_}i}-\texttt{mod{\_}i}$>0.075$ (referred below as \textit{magnitude} criterion), or $\log_{10}\mathcal{L}_{star,i}<-11$ (resp. \textit{stellarity} criterion).
 The \texttt{psf} and \texttt{model} magnitudes, as well as the $i-$band stellarity likelihood $\log\mathcal{L}_{star,i}=$\texttt{lnLStar{\_}i}, are taken directly from the SDSS catalog\footnote{A description can be found in the SDSS Schema Browser, e.g. at \texttt{http://skyserver.sdss.org/dr7/en/help/browser/browser.asp}, Table `PhotoObjAll'.}. Their definition is, in fact, somewhat survey-specific and different thresholds must be explored for ATLAS, DES and PS1, separately for each survey.
Since this search is tailored on lensed quasars, I further select objects satisfying
\begin{eqnarray}
\nonumber W1-W2>0.55,\  2.2<W2-W3<3.8\\
\nonumber W1<17.0, W2<15.4, W3<11.6, \\
\nonumber \delta W1<0.25,\ \delta W2<0.3,\ \delta W3<0.35,\\
\nonumber 2.2<i-W1<5.75,\ i-W3<8.9\\
\nonumber g-i<\mathrm{max}(0.65; 1.2(i-W1)-2.4)\\
g-i<2.55,\ r-z<1.5,\ 15.0<i<20.5
\label{eq:colcuts}
\end{eqnarray}
which eliminates most stellar contaminants, blue galaxies and low-redshift quasars. Here, $griz$ magnitudes are SDSS \texttt{model} magnitudes, WISE $W$`X' magnitudes are \texttt{wXmpro} and $\delta W$`X' are the uncertainties \texttt{wXsigmpro} on the corresponding magnitude (X=1,2,3). Magnitude-selected extended objects will also be split into `\texttt{c0}' objects, having
\begin{equation}
 W2-W3<\mathrm{max}[2.7;\, 3.15+1.5(W1-W2-1.075)],
\label{eq:wisecols}
\end{equation}
and `\texttt{c1}' objects, occupying the remaining wedge in WISE colours. This colour distinction delimits the locus where all SQLS training-set objects lie, and roughly traces the distinction between quasar-dominated and galaxy-dominated objects \citep[e.g.][]{wri10}.

Main contaminant classes are identified using Gaussian population mixture as a clustering algorithm, which has been validated as a means of object classification across different surveys \citep[e.g.][where hybrid cuts on $g-i$ and $i-W1$ similar to the above have been tested]{bov11,dip15,che16,wil17,tie17}. A recurring theme of semi-supervised clustering and classification methods is whether class parameters should be initialized: (\textsc{i}) based on where we expect different objects to lie \textit{a priori};  or (\textsc{ii}) identifying clusters independently and labeling them \textit{a posteriori}. Here, I choose the second option, even though some clusters can be easily interpreted in terms of known object populations, as will be done below.

Each class is then described by a mean $\boldsymbol{\mu}$ and a covariance matrix $\boldsymbol{\Sigma}.$ The relative class abundances will not be used at classification stage. To each object, with feature vector $\mathbf{f}_{i},$ a pseudo-distance to the $k-$th class is defined as
\begin{equation}
d_{i,k}=\langle(\mathbf{f}_{i}-\boldsymbol{\mu}_{k}),\boldsymbol{\Sigma}_{k}^{-1}(\mathbf{f}_{i}-\boldsymbol{\mu}_{k})\rangle/2
\end{equation} 

The mean and covariance of each class are computed iteratively using Expectation-Maximization. In order to ensure convergence, I borrow from the strategy of adaptive second moments \citep{ber02}: at each step in the computations of $\boldsymbol{\mu}_{k}$ and $\boldsymbol{\Sigma}_{k},$ each object is additionally weighed with $\mathrm{e}^{-d_{i,k}},$ multiplying the covariances by 2. This avoids iteration instabilities due to neighbouring classes or `large' widths in the classes. Some of the mean features are kept fixed, as they have values outside those of the queried sample (e.g. $W1-W2$ of galaxies) or are unstable to Expectation-Maximization.

The first four classes are common to objects selected via magnitude or stellarity morphological criteria. Most of the magnitude-selected sample is described by six clusters, whereas eight clusters are needed to encompass the majority of the stellarity-selected sample. Table~\ref{tab:feattab} lists the averages $\boldsymbol{\mu}_{k}$ and for the sake of  brevity the widths $\sqrt{\Sigma_{k,jj}}$ ($j=1,...,7$) of the classes. By comparing the colours of different classes with those from spectroscopic subsamples  \citep[see e.g.][]{wil17}, one can roughly associate the first four classes with: (\textsc{i}) isolated quasars at $z_{s}\lesssim 0.35,$ which have bright $W2,$ high $i-W1$ and low $g-i$ but can extend to redder colours due to contribution from their host galaxies; (\textsc{ii}) isolated quasars at redshift $z_{s}\approx2,$ with fainter $W2,$ $g-i\lesssim0.6$ and lower $i-W1;$ (\textsc{iii}) narrow-line galaxies at $z\approx0.2-0.3,$ with $W2$ and $i-W1$ comparable to $z_{s}\approx2$ quasars but higher $g-i;$ (\textsc{iv}) and fainter galaxies with 
 $W1-W2$ spanning a wide range. In fact, the empirical WISE colour-magnitude cuts of  \citet{ass13} were designed to minimize contamination from the fourth class leaking into WISE-selected quasar samples.
\begin{figure*}
 \centering
 \includegraphics[width=0.3\textwidth]{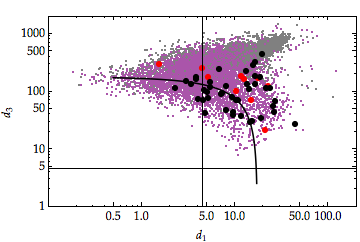}
 \includegraphics[width=0.3\textwidth]{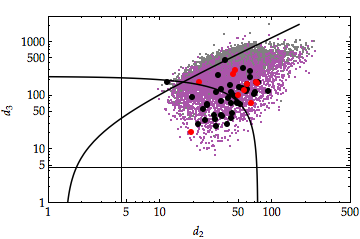}
 \includegraphics[width=0.3\textwidth]{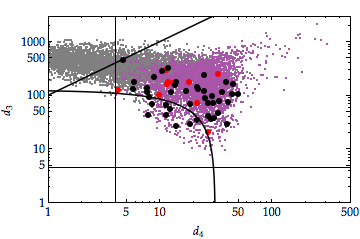}\\
 \includegraphics[width=0.3\textwidth]{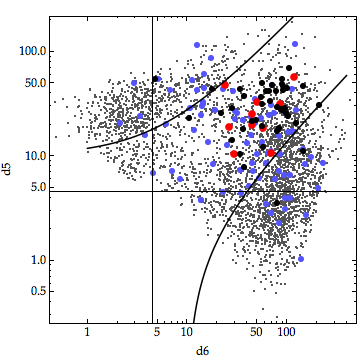}
 \includegraphics[width=0.3\textwidth]{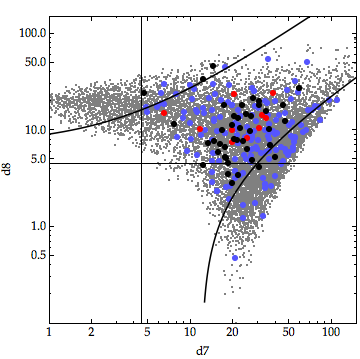}
 \includegraphics[width=0.3\textwidth]{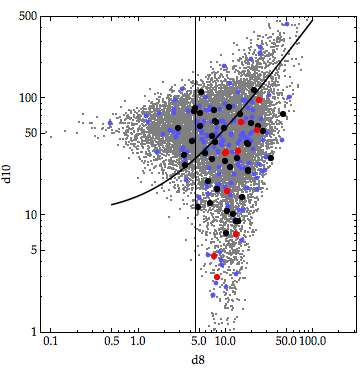}\\
 
\caption{{Cuts in pseudo-distances from main clusters/classes. Red (resp.) black bullets mark SQLS lenses (resp. false positives) in the training set.
 \textit{Top:} first set of cuts on $d_1,...,d_4;$ grey (resp. purple) swarms trace magnitude-selected (resp. stellarity-selected) objects \textit{Bottom:} cuts in $d_5,...,d_{10}$ depending on the extended-morphology criterion; light-blue bullets mark the candidates selected after visual inspection of the first-pass targets, and \texttt{c0} (resp. \texttt{c1}) targets correspond to high $d_6$ (resp $d_5$). The axes are shown mostly as a guidance to the eye.
 }}
\label{fig:cutplots}
\end{figure*}

\begin{figure}
 \centering
 \includegraphics[width=0.45\textwidth]{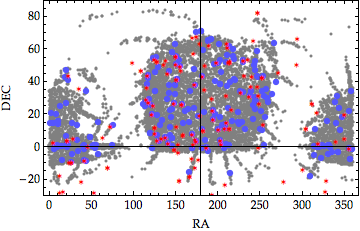}\\
 \includegraphics[width=0.45\textwidth]{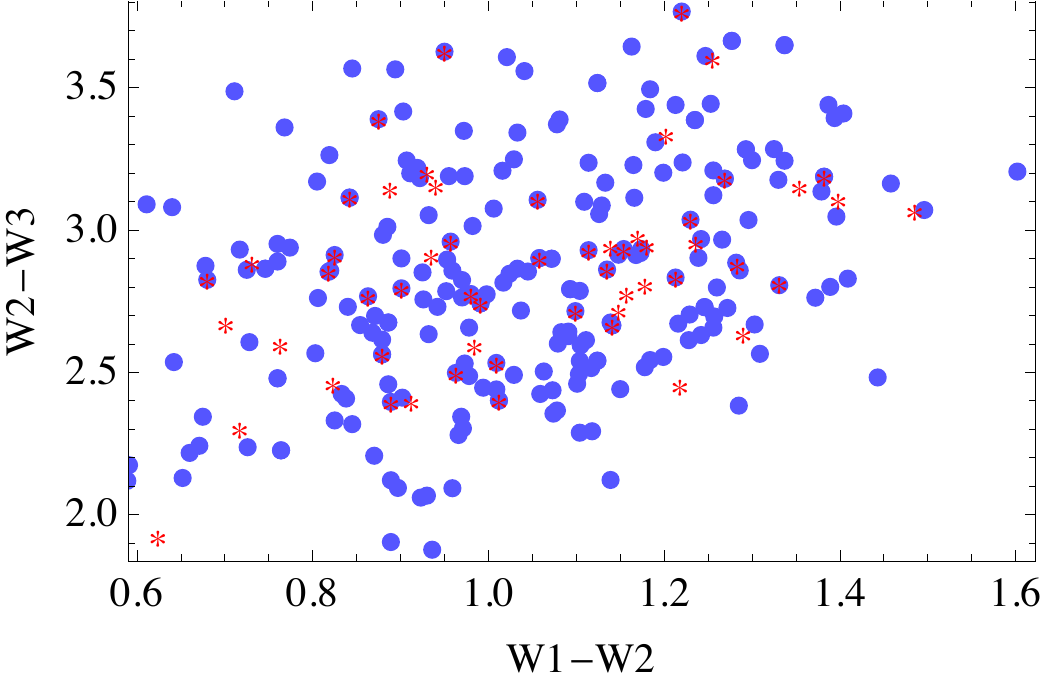}\\
 \includegraphics[width=0.45\textwidth]{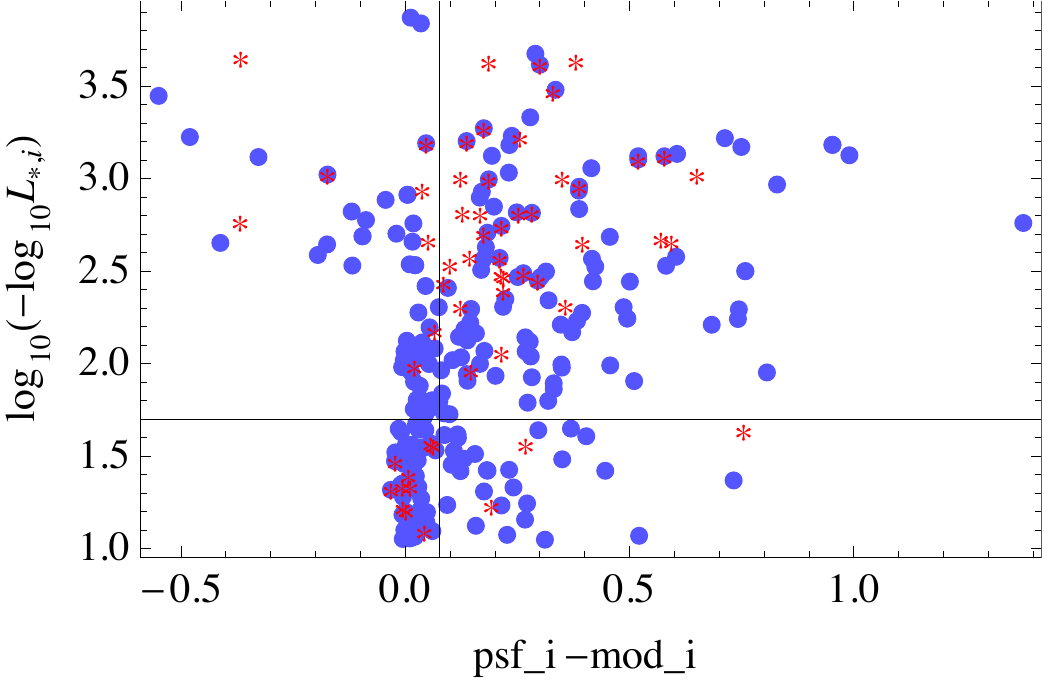}\\
\caption{{Distribution of query-selected objects (grey), candidates (light-blue bullets) and known lenses (red stars). The SDSS coverage is not homogeneous, as reflected at all stages of query, target- and candidate-selection and by the known systems, which have consistent distributions in WISE colours and stellarity/\texttt{psf-mod} distribution. The axes are shown mostly as a guidance to the eye.
 }}
\label{fig:canddist}
\end{figure}

\subsection{Adapting on SQLS Training Set}
Once the classes/clusters are defined, targets must be selected based on how far they lie from different classes. In order to do so, I use linear combinations of the pseudo-distances $d_{k}.$
The first set of pseudo-distance cuts, using $d_{1},...,d_{4}$ as
\begin{eqnarray}
\nonumber d_1\geq 4.5, d2\geq 15, d_4\geq 4, \\
\nonumber 12d_2-17\geq d_3\geq 225-3d_2\\
\nonumber d_3\geq 175-10d_1, d_3\geq 6d_1-20\\
125-4d_4< d_3\leq 100d_4
\label{eq:firstcuts}
\end{eqnarray}
is displayed in Figure~1. Some of the cuts can be made more or less restrictive, with the aim of retaining as many quasar-like systems as possible, while rejecting galaxies and SQLS false-positives without penalizing the SQLS true-positives.

With this combination, nine out of ten training-set lenses are retained, and twenty out of forty false-positives are rejected.
At face value, this would mean that outlier selection is 90$\%$ complete with respect to the SQLS and has half as many contaminants, but matters are more complicated. One the one hand, the SQLS and BQLS relied on spectroscopic information, so contamination by narrow-line galaxies is completely suppressed there, while it is non-negligible in photometric searches. On the other, the same spectroscopic pre-selection limits the SQLS and BQLS to objects for which spectra are available, and has in fact precluded the discovery of some noteworthy lenses.

When the first cuts of equations~(\ref{eq:colcuts}, \ref{eq:firstcuts}) are applied to objects in the SDSS-DR12 footprint, {8543} (resp. 15839) objects are still left in the magnitude-selected (resp. stellarity-selected) sample\footnote{These are \textit{non-unique} identifications, as some of the objects are identified and retained multiple times, through the whole selection process from query to final candidate sample. Non-repeated objects are $\approx60\%$.}.
In the colour-colour diagrams of remaining objects, there are some overdensities corresponding to (apparently) isolated quasars with $z_{s}=0.5\pm0.1,$ based on the available fibre spectra. Their $\boldsymbol{\mu}$ values are different between magnitude-selected (two clusters, $k=5,6$) and stellarity-selected (four clusters, $k=7,...,10$) objects. The lower panels of Figure 1 show where objects surviving the cuts of equation~(\ref{eq:firstcuts}) lie in terms of the distances from additional clusters ($d_{5},...,d_{10}$).

To further guide the selection, I visually inspected the first-pass targets and retained 82 magnitude-selected and 157 stellarity-selected candidates, which are not `obvious' kinds of contaminants such as blue galaxies or nearby quasars and Seyfert galaxies. Their pseudo-distance distribution, occupying mostly the outskirts of the object clusters (see fig.~1), suggests the following cuts
\begin{equation}
0.15d_6 -1.5\leq d_5\leq 1.75d_6+10
\label{eq:magscuts}
\end{equation}
for magnitude-selected objects, and
\begin{eqnarray}
\nonumber 2.0d_7+7\geq d_8\geq 0.25d_7 -3\\
d_{10}\leq 4.5d_8+10,\ d_9\leq 10d_8+10
\label{eq:stescuts}
\end{eqnarray}
for stellarity-selected objects. Once applied to the first-pass candidates, they result in 3728 magnitude-selected and 4712 stellarity-selected targets, $60\%$ of which are non-repeated catalog entries. This has reduced the initial $\approx5\times10^5$ queried objects to a manageable sample for visual inspection; in fact, the cuts in $d_1,...,d_4$ were already enough to obtain a reasonable reduction in objects to be inspected, whence the 82+157 candidates were obtained.

\subsection{Blind Test on SDSS}
Using the CASTLES+SQLS \textit{test} set introduced above, we can quantify how many known lenses are lost at each stage and why. At pre-selection level, 57 of the 132 test objects are retained, due primarily to the extendedness criteria as noticed already by \citet{wil17}, and secondarily to WISE colour selection. Of these, 35 remain after the first cuts (in eq.~\ref{eq:firstcuts}); 36 satisfy equations~(\ref{eq:magscuts},\ref{eq:stescuts}) and only 22 satisfy the cuts in all pseudo-distances. Most of the rejected lenses lie close to the selection boundaries, which could be re-adjusted \textit{post hoc} to increase the completeness. However, I preferred to perform a blind test of this method, trained solely on 10 lenses and 40 non-lenses.

Figure~\ref{fig:canddist} shows the distribution of queried objects (grey swarm), candidates selected after $d_1,...,d_4$ cuts (blue bullets), and known lenses (red star symbols). One every 40 queried objects is shown, for convenience. Inhomogeneous spatial coverage is a direct consequence of the SDSS scanning, and affects the distribution of targets, candidates and known lenses. Candidates and known lenses have compatible distributions in WISE colours (not used by previous campaigns) and morphological parameters.

This outlier-selection seems to retain known quads and doubles alike. Some lenses are present multiple times in the queried sample, the most popular being: J2343-0050 (78 matches!), J1001+5553 (20 m.), J0145-0945 (7 m.), J1206+4332 (6 m.), J0806+2006 (5 m.), J1304+2001 (4 m.). Besides these, 10 lenses are flagged 3 times by the object query, and 20 are flagged twice. Some BQLS objects, not included in the test set, have been rediscovered as well.
When applied to the DES catalog, with suitably translated magnitudes \citep[as discussed by][]{agn17a}, it recognized both large-separation lenses like DES0408 \citep{lin17} and small-separation blends like DES0115 \citep{agn15b}. The main reason is that, while all these lenses have markedly quasar-like WISE colours, their hybrid colours are not typical of unlensed and low-redshift quasars.

Whether quasar `pairs', which are interesting for other applications, are eliminated or retained depends chiefly on how much their overall colours resemble those of low-redshift quasars. Based on the results of different spectroscopic follow-up campaigns\footnote{Papers in preparation by Williams et al., Agnello et al., and by the STRIDES collaboration (\texttt{strides.astro.ucla.edu}).}, quasar pairs with sources at $z_{s}\approx2$ are present among quasar lens candidates. Some quasar pairs with nearly identical spectra are also present, but whether they are lenses or physically associated quasars will require deeper follow-up.

\section{GAIA multiplet detection}

By using different all-sky surveys, one can combine their separate advantages and exploit the largest possible footprint. In particular: WISE enables a mid-IR selection of quasar-like objects, gathering the light from different quasar images within the same $\approx6^{\prime\prime}$ beam; while GAIA offers higher spatial resolution, enabling a distinction between isolated quasars and quasars with companions within the same WISE beam. From this viewpoint, combining WISE and GAIA for quasar lens searches is reminiscent of the CLASS/JVLA strategy of radio-loud detection and higher-resolution follow-up.

Here, I illustrate the general properties of objects selected with WISE $W1,W2,W3$ and GAIA DR1 data, using the CASTLES+SQLS known systems as guidance. With the addition of SQLS quasar pairs, the test-set amounts to 197 objects. Visual inspection is needed to obtain a candidate sample, so I will explore the performance of this combined search over three footprints: one covering the SDSS-DR12; one covering the ATLAS-DR3; and one covering slightly more than the DES-Y3. For SDSS and ATLAS, I will also rely on public survey catalogs and images, whereas for the DES footprint I will delineate some general properties based solely on WISE and GAIA data. To quantify how many lens candidates can be expected with this method, I select objects with $W1,$ $W2,$ $W3$ as in Section~2 (over the three considered footprints) and match them to GAIA using a 6$^{\prime\prime}$ search radius. In the SDSS and ATLAS, when $griz$ catalog magnitudes are available, I also require that objects satisfy the colour-magnitude requirements of eq.~(\ref{eq:colcuts},\ref{eq:wisecols}). As a separate footprint, to examine the role of hybrid optical-IR colours, I will study objects with dec$>-30$deg that are detected also in the iPTF catalog. These have the advantage that they can also be visually inspected using PS1 images, and that quasar-like objects can be further selected based on their variability signatures \citep{sch12}.

As an illustration of the method, and to minimize contamination by star-quasar pairs, the WISE-GAIA matches will be performed on objects with $G>15.0$ and $\left| b\right|>20.0$ and visually inspected for $\left| b\right|>30.0$ when possible. This is also why the recent quad candidate of \citet{ber17} is discarded at preselection, while (at least) four known quads are retained: HE0435 \citep{wis02}, RXJ1131 \citep{slu03}, PG1115 and SDSS~J1433+60.

\begin{figure}
 \centering
 \includegraphics[width=0.45\textwidth]{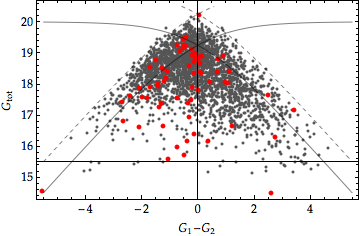}\\
 \includegraphics[width=0.45\textwidth]{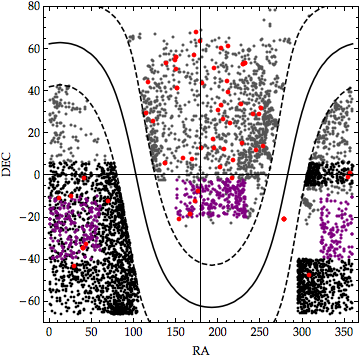}
 \includegraphics[width=0.45\textwidth]{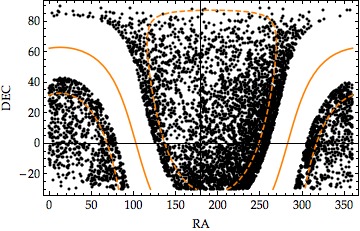}\\
\caption{{\textit{Top:} total magnitude versus magnitude difference of individually resolved components, for WISE objects split by GAIA. The lines delimit systems with a faintest image with $G=20$ (full) or $G=21.$ (dashed). \textit{Middle:} Spatial distribution of WISE-GAIA multiplets Grey (resp. purple, black) points correspond to the SDSS (resp. ATLAS-DR3, pseudo-DES) footprint. In both panels, red bullets indicate known quasar lenses/pairs. \textit{Bottom:} WISE-GAIA multiplets with dec$>-30$; the dashed contours correspond to $b=\pm30$deg.
 }}
\label{fig:gaiadist}
\end{figure}
\subsection{General behaviour}

Most known lenses and pairs correspond to one detection per system in WISE, whereas 55 of them are recognized as separate objects in GAIA-DR1. Their overall properties are summarized in Figure~\ref{fig:gaiadist}: the WISE-GAIA match can be performed down to $G=20.7;$ most lenses with a GAIA multiplet counterpart have $G<20;$ and the flux-ratios of different components are not extreme, mostly $\left|2.5\log_{10}(f_1/f_2)\right|\lesssim2.$

The grey points in Figure~\ref{fig:gaiadist} display general properties of the WISE-GAIA multiplets, for a random sub-set, showing the same behaviour as for the known lenses in red. This first match results in 1868 multiplets in SDSS-DR12, 648 in ATLAS-DR3, and 2679 in the approximate DES footprint. All of these are manageable numbers for a quick visual inspection, and have led to new lenses being discovered with this method, which will be reported elsewhere.

Exploiting the homogeneous (sometimes spectroscopic) coverage of SDSS, we can further quantify the fraction of promising multiplets. A quick visual inspection reveals different contaminant classes: galaxy groups and mergers; stars, especially at low galactic latitudes $\left|b\right|<33$deg; quasar-star and quasar-galaxy LOS alignments. Those that are not `obvious' contaminants amount to 127 candidates, some more convincing than others. By construction, they include known lenses with different configurations (quads and doubles) and image separations.

The situation in other footprints is complicated by the patchy (ATLAS) or limited (DES) coverage of public catalogs. Visual inspection of ATLAS-DR3 targets with at least one valid magnitude results in 28 unique candidates. The known quad RXJ1131 is flagged four times by this search (corresponding to its four quasar images), even though it is covered just in $i-$band in the public footprint. Visual inspection produces one candidate every 20-30 targets, so one can expect 60-100 candidates to be selected in the DES footprint with this method; the depth, image-quality and $grizY$ coverage of DES also help to further select the most promising systems for follow-up.

Since the multiplet search targets objects with quasar-like WISE colours and nearby neighbours, it also retains line-of-sight alignments of quasars and stars. This aspect is further discussed below.

\subsection{Hybrid Colours}
\begin{figure}
 \centering
 \includegraphics[width=0.45\textwidth]{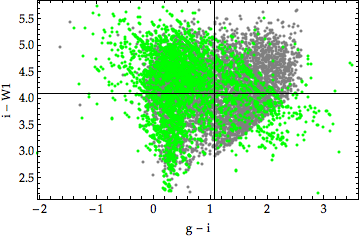}\\
 \includegraphics[width=0.44\textwidth]{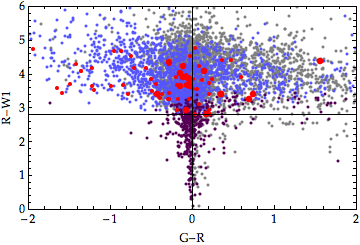}\\
 \includegraphics[width=0.45\textwidth]{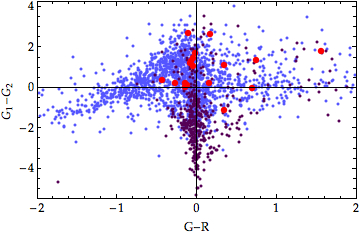}\\
 \includegraphics[width=0.44\textwidth]{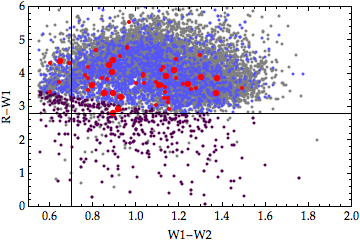}
\caption{{Using hybrid colours across three surveys, for patchy ground-based coverage. \textit{Top:} Grey (resp. green) points show $g_{\rm sdss}-i_{\rm sdss}$ (resp $0.175+G-i_{\rm sdss}$) vs $i_{\rm sdss}-W1$ for SDSS queried objects (resp. GAIA multiplet counterparts). \textit{Lower} panels: hibrid colours of WISE-GAIA multiplets that also fall in the iPTF footprint, with purple (resp. blue) indicating objects with lower (resp. higher) $R-W1;$ red bullets indicate known lenses.
 }}
\label{fig:gaiacols}
\end{figure}

Since hybrid colours help separate quasar-like objects from galaxies and stars, an ideal search would incorporate the multiplet information from GAIA with optical/IR colours to reduce contamination. 
The top panel in figure~\ref{fig:gaiacols}, displaying hybrid colours with GAIA, SDSS and WISE, suggests that these may still be used to aid the WISE-GAIA multiplet detection in the case of partial (e.g. only $i-$band) ground-based coverage. There, I am using $0.175+G-i$ (green points) instead of $g-i$ (grey swarm) to quantify the red excess of SDSS objects corresponding to GAIA multiplets. Indeed, most quasar-like objects have compatible colours across SDSS and GAIA. Some objects have $G-i$ vs $i-W1$ falling beyond the pre-selection boundaries, and may be LOS pairs of quasars and stars that are resolved by GAIA but not by the SDSS.

A similar scenario is shown by matching WISE-GAIA objects (singlets and multiplets) with the iPTF. The lower panels in fig.~\ref{fig:gaiacols} shows their hybrid colours across the three surveys, with $R-$band from the iPTF when available. Similarly to the pseudo-DES footprint, objects were selected based only on WISE colours, so a large contamination by LOS quasar-star pairs is present, as seen in the spatial distribution of multiplets: many abound at $\left| b\right|<30$deg, besides clumps corresponding to M31 and M33. In terms of colours, flux ratios $G_{1}-G_{2}=2.5\log_{10}(f_{1}/f_{2})$ and separations, different object classes can be recognized: (\textsc{i}) quasar-like objects, with $R-W1\gtrsim2.8,$ $R-W2\gtrsim3.9,$ themselves divided in three main clusters at $W1-W2=(0.750\pm0.125),$ $(0.94\pm0.07),$ $(1.225\pm0.175)$; (\textsc{ii}) quasar-star pairs in which the star dominates in the optical, with low iPTF-WISE colours and $G-R\approx -0.02\pm0.05,$ with object separations distributed smoothly and regardless of colours; (\textsc{iii}) `bluer' objects, with $G-R\lesssim-0.2,$ $\left| G_1-G_2\right|\lesssim2.0$, and typical separations $\gtrsim5.0^{\prime\prime}.$

Different object classes are all intersecting, at least in projection, and again the known quasar lenses and pairs seem to occupy the outskirts of different populations. Quasar-star objects with low $R-W1$ seem just the visible `tip of the contaminant iceberg', since multiplets with $\left| b\right|<30$deg (perhaps mostly quasar-star pairs) form a tight sequence with $G-R\approx -0.02\pm0.05$ but extend to higher values of $R-W1,$ depending on the relative contribution of quasar and star. Since quasars dominate in the infrared, the WISE $W1-W2,$ $W2-W3$ colours or quasar-star pairs are indistinguishable from those of isolated and lensed quasars.

The analysis of hybrid colours suggests that an integrated approach, incorporating both the multiplet detection and outlier selection, may yield cleaner lens candidate samples. This has not been used for this work yet, and is shown as a possible extension for future investigation. It may be replaced altogether by GAIA colours, based on $G_{\rm BP}$ and $G_{\rm RP},$ if these are available in future releases.

\begin{figure}
 \centering
 \includegraphics[width=0.45\textwidth]{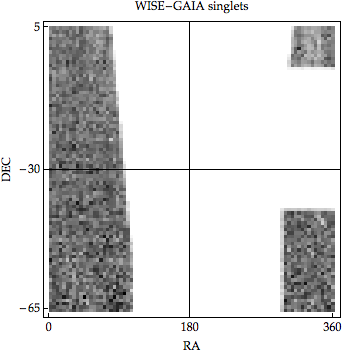}\\
 \includegraphics[width=0.44\textwidth]{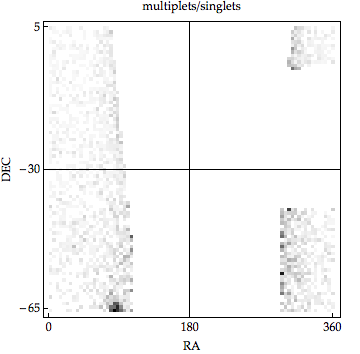}\\
\caption{{Distribution of (non-duplicate) WISE-GAIA singlets and multiplets over the pseudo-DES footprint. \textit{Top:} density of single matches; \textit{bottom:} ratio of multiplets to singlets, showing the same inhomogeneities identified from fig.~\ref{fig:canddist}. The edge of the MW disk is visible in the increased multiplet counts at low $\left| b\right|.$
 }}
\label{fig:mulsing}
\end{figure}
\subsection{Foregrounds: Galactic Substructure}
Where the stellar density is higher, e.g. at low Galactic latitudes, the WISE+GAIA multiplet search triggers more often. While this is a nuisance for quasar lens searches, it also means that it can be used to locate stellar overdensities such as satellite galaxies, globular clusters and streams. Figure~\ref{fig:gaiadist}, plotting multiplets among \textit{non}-duplicate GAIA detections, shows some of these.
Besides the Large Magellanic Cloud and the Orphan Stream, some Southern candidate streams appear. I designate them as WG1, with r.a.,dec.$\approx$(64.16,-23.2) to (97.2,-55.5); WG2, with r.a.,dec.$\approx$(70.5,-51.4) to (95.7,-47.1); WG3, with r.a.,dec.$\approx$(304.1,-47.7) to (351.3,-44.8); and WG4, with r.a.,dec.$\approx$(300.0,-49.4) to (333.5,-66.0).
Of these, two (WG3,WG4) may have counterparts independently identified\footnote{$^\ddag$ E.~Balbinot, private communication.}$^{\ddag}$ in DES data (Shipp \& DES Collaboration, 2017, in prep.). 
The sharpest stream is WG1, WG2 is weaker and barely visible in the DES search$^{\ddag}$, and the thin stream WG3 seems to cross the Galactic plane, whereas WG4 is considerably thicker and may be a superposition of two, almost-parallel streams. Their quoted endpoints should be solely a guidance to the eye: they are not the primary aim of the WISE-GAIA multiplet search and so they are roughly traced by the detected quasar+star alignments. Besides WG1,...,WG4, the thin stream of \citet{bal16} can be seen crossing the ATLAS and DES footprints.

Other candidate streams can be seen at low grazing angles from the Galactic plane, or in the ATLAS footprint, but are less sharp and may be given by patchy footprint coverage, as traced by the distribution of WISE-GAIA quasar-like singlets. An example of spurious overdensity, in the Northern Galactic Hemisphere, is given by two wide regions with low WISE-GAIA singlet density separated by a thin strip that has more complete coverage, which could otherwise be confused for a stream extending from ra,dec$\approx(160,15)$deg to ra,dec$\approx(150,45)$deg. The four candidates WG1,...,WG4 seem more robust than others because they do not seemingly coincide with inhomogeneities in the WISE-GAIA coverage: as shown by fig.~\ref{fig:mulsing} in the ratio of multiplets to singlets, the same overdensities corresponding to WG1,...,WG4 remain once their distribution is normalized to that of singlets.

Given the primary scope of this paper, i.e. lensed quasar searches, the identification of WG1,...,WG4 may be regarded as an unintended by-product. Searches for Galactic substructure via stellar overdensities have been applied extensively in the past to the SDSS \citep{bel06,zuc06,bel07,irw07,kop07,bel09}, ATLAS \citep{bel14} and DES \citep{bec15,kop15,bal16,li16} footprints, and have also led to the discovery of two star clusters using GAIA data alone \citep{kop17}. The search outlined here differs from those as it targets neighbours within the same $6^\ase-10^\ase$ radius, as opposed to overdensities on arc-minute scales from binned star counts.

Two steps forward may be envisioned, enabling an all-sky WISE+GAIA search of Galactic substructure down to the $G=20.7$ detection limit. First, it may be tailored directly on stellar objects ($W1-W2\lesssim 0.3,$ instead of $W1-W2>0.55$), and may also be augmented with 2MASS \citep{skr06} information, such as $J-W1<1.2$ \citep{vic16}. The quasar-like colour selection and match to GAIA are, in fact, one reason for the circular gaps in the distribution of multiplets over the sky, due to crowding or higher extinction. Second, while WG1,...,WG4 were identified solely based on visual inspection, a more complete search may automatically select associations of multiplets that lie on the same line over $\approx5-10$ degree-long distances. This would also help identify more diffuse streams from more massive projenitors, which \citep{bg17} can otherwise be confused with random fluctuations in the distribution of halo stars.

Even fainter magnitudes may be reached by trading the resolution of GAIA with deeper $grizY$ magnitudes from DES in the South or PS1 in the North. In this case, the WISE pre-selection might be dropped altogether, by-passing completeness limits at $i\gtrsim21;$ however, a quasar-like WISE-GAIA \textit{singlet} selection is still useful to characterize the survey coverage, and hence evaluate spurious substructure detections.

\section{Conclusions and future prospects}
I have introduced two new methods to find quasar lenses, and illustrated their performance on publicly-accessible footprints, using {training}- and {test}-sets of known lenses and pairs in the CASTLES and SQLS databases. The first concentrates on objects that are not well described as isolated quasars or blue galaxies; the second joins WISE and GAIA to detect multiplets of point-sources with overall quasar mid-IR colours. Besides recovering known lenses (with varying degree of success), both searches have led to new discoveries, reported elsewhere. A by-product of the WISE+GAIA search has been the identification of Galactic stream candidates that (as I later learned) are being found within the DES collaboration using different techniques.

A common feature of many searches, including the outlier-selection and multiplet-detection, is the trade-off between completeness and manageable size of candidate samples to be visually inspected and followed-up at the telescope. Searches concentrating on `lens' colours \citep[e.g. this outlier-selection and the population-mixture classification by ][]{wil17} discard a high fraction of lenses with colours akin to nearby unlensed quasars. This is a price to pay in order to find systems with sources at higher redhift, or that could not be pre-selected for SDSS fibre spectroscopy due to their unusual colours. On the other hand, searches tailored on `quasar' colours, amending the loss of lenses with quasar-like colours, result in large candidate samples to be examined: according to \citet{ost17}, a double was found after eyeballing 5000 objects brighter than $i=$19 in a 1500deg$^2$ footprint. By comparison, the outlier selection resulted in 5064 SDSS targets brighter than $i=20$.

The pseudo-distance cuts reduce the number of objects considerably, from $5\times10^5$ to $\approx8600$ targets, of which $\approx60\%$ non-repeated objects, down to 239 final candidates, to the price of a harsh selection on known lenses. Even though few systems are rejected by each pair-wise combination of pseudo-distances, and most of the rejected lenses lie close to the cut boundaries, all cuts combined reduce the overall completeness to $30\%,$ or $50\%$ when only cuts in $d_{1},...,d_{4}$ are used. Since the number of targets for visual inspection is manageable, one could loosen the cuts in $d_5,...,d_{10};$ this would mean that the whole test-set would now be a training-set for the method. Despite the $50\%$ completeness with respect to already-known lenses, previously unknown lenses have been discovered with this technique \citep[e.g. the new quad J1433+60,][]{agn17b}, which suggests that outlier-selection is indeed complementary to previous searches.

Comparable numbers, from query to final candidates, are found when applying the WISE-GAIA search to the SDSS footprint. The resolving power of GAIA enables multiplet detection for objects that are otherwise blended by the pipelines of ground-based surveys. New lenses, not found by other techniques, have been discovered with this method and will be presented elsewhere. Its performance, however, is not obvious to characterize: while it recognizes the two quasar images of the close double DES~J0115 \citep{agn15b}, and all quasar images of the fold PG1115 \citep{wey80}, it detects only two images in the DES~J0408 quad \citep{lin17}, and two the WFI2033 quad \citep{mor04}, whereas the WFI2026 quad is seemingly unresolved. Due to $G=20.7$ depth and pipeline-specific choices in resolving close pairs with high flux-ratios, roughly one every three known lenses/pairs is resolved by GAIA.

Hybrid colours may be used as additional information, either using (limited) ground-based information or via $G_{\rm RP},$ $G_{\rm BP}$ photometry, if available in future releases. This will be particularly interesting in view of the EUCLID mission, which should provide a wide-field, deep, and sharp  counterpart to the slitless spectroscopic Hamburg-ESO survey \citep{wis00}, from which some remarkable quasar lenses have been identified.

Classification via clustering and outlier-selection has been implemented with different methods, for other purposes, over the last two years. Clustering in colour-space has been used to evaluate photometric redshifts \citep{rah16a,rah16b}, select LRGs \citep{roz16} and identify galaxy clusters from improved red-sequence finders \citep{ryk14}. Outlier-selection, implemented via random-forest classification on SDSS fibre spectra of galaxies, has been used to select peculiar galaxies of various types \citep{bar17}. While the search in this paper was tailored on (lensed) quasars, and others were concentrated on galaxies, there is significant scope for integrating these into a comprehensive photometric classification across wide-field surveys and selection of the most peculiar objects.

A by-product of the WISE-GAIA multiplet search is the detection of (candidate) Milky Way streams, as coherent structures traced by stars in front of WISE quasar candidates. Multi-band follow-up is needed, however, to confirm them and characterize their stellar populations. A WISE-GAIA search can be tailored on streams, using a different WISE selection and possibly augmenting with near-IR colours to separate extra-galactic objects.

\section*{Acknowledgments}
I wish to thank Eduardo Balbinot, Nora Shipp, Alex Drlica-Wagner for the comparison between my MW stream candidates and theirs; and Elizabeth Buckley-Geer, Huan Lin, Paul Schechter and Tommaso Treu for discussions on quasar lens searches. Careful comments by the anonymous referee have led to clarifications and improvements of this paper. I gratefully acknowledge hospitality by the ITC-Harvard, where this work was completed.

\label{lastpage}
\end{document}